\documentclass[pra,showpacs]{revtex4}
\usepackage{graphicx}

\begin{document}
\title{Entanglement dynamics for two interacting spins}
\author{Marcel Novaes}

\affiliation{Instituto de F\'{i}sica ``Gleb Wataghin",
Universidade Estadual de Campinas, 13083-970 Campinas-SP, Brazil}

\begin{abstract}
We study the dynamical generation of entanglement for a quantum
system consisting in a pair of interacting spins, $s_1$ and $s_2$,
in a constant magnetic field. Three different situations are
considered: a) $s_1=s_2=1/2$, b) $s_1\to \infty, s_2=1/2$ and c)
$s_1=s_2\to\infty$, corresponding respectively to a fully quantum
system (two interacting qubits), a quantum degree of freedom
coupled to a semiclassical one (a qubit in contact with an
environment) and a fully semiclassical system, which displays
chaotic behavior. The time evolution of entanglement for initially
separable states is measured using the von-Neumann entropy and the
linear entropy, and a comparison is made between the two.
Compactness of the phase space and chaos in the classical limit
are seen to have a strong influence on the recoherences of the
system.

\end{abstract}


\maketitle

Recent technological advances have made it possible to create and
manipulate individual quantum states in the laboratory, thus
allowing direct observations of entanglement and decoherence
\cite{prl77mb1996,n404cas2000,rmp73jmr2001,n413bj2001,n423jwp2003,n428bbb2004,n429mr2004,n429mdb2004},
concepts that are central to quantum computation and quantum
information \cite{nielsen2000,roadmap}. Two subsystems $1$ and $2$
are said to be entangled if it is not possible to write the total
system density matrix in the form
\begin{equation}
\rho=\sum_i p_i \rho_1^i\otimes\rho_2^i,
\end{equation}
and decoherence is the suppression of off-diagonal elements in a
subsystem reduced density matrix \cite{rmp75whz2003}. Quantifying
entanglement, however, is a difficult task. Many different
measures have been proposed, such as entropy of formation, entropy
of distillation, and many different distances
\cite{prl78vv1997,pra57vv1998,rmp74vv2002,jpa36je2003,prl92mhp2004}.
All of them are very hard to calculate in practice, although
Wootters \cite{prl80wkw1998} has provided a closed form for the
entropy of formation in the case of a system consisting of only
two qubits. If the initial state is separable, the entropy of
formation for a bipartite system reduces to the von-Neumann
entropy (we reserve the letter $S$ for spins)
\begin{equation}
\delta_N=-{\rm Tr} \left[\rho_2\log_d\rho_2\right],
\end{equation}
where $\rho_2$ is the reduced density matrix of the subsystem $2$
and $d$ is the dimension of its Hilbert space. Another
easy-to-handle measure is the linear entropy (or idempotency
defect)
\begin{equation}
\delta=\frac{d}{d-1}(1-{\rm Tr} \left[(\rho_2)^2\right]).
\end{equation}
When one considers only globally pure states (as we do here), both
the linear and the von-Neumann entropy are $0$ for separable
states and $1$ for maximally entangled states. If the initial
state is separable, one is in fact considering a dynamical,
deterministic, generation of entanglement. In this situation how
fast entropy grows indicates how fast one subsystem suffers
decoherence due to entanglement with the other.

Another interesting question is about the dependence of quantum
entanglement upon the classical dynamics of the system, more
specifically what is the role of chaos in the decoherence process.
This has been the subject of many recent investigations. In a
seminal work Furuya {\it et al} \cite{prl80kf1998} have studied
the Jaynes-Cummings model without the rotating-wave approximation,
and found the entanglement rate to be greater for chaotic initial
conditions. In later works \cite{pre60rma1999,pra64rma2001} they
have shown that a regular initial condition can sometimes lead to
faster entanglement than a chaotic one and that recoherences were
related to the shape of the spin orbits. In the past few years
much work has been done in this area, considering continuous
variables \cite{jpa36ls2003,pa338rma2004}, tops
\cite{pre60pam1999,pre67hf2003,jpa36mz2003,pre69jnb2004}, and spin
chains \cite{prl91sm2003,pra69lfs2004,lages2004}.

In the present work we consider two spins, $S_1$ and $S_2$, in a
constant magnetic field $\vec{B}=B_0 \hat{z}$, interacting
according to the Hamiltonian
\begin{equation}
H=\epsilon_1 B_0 S^z_1 +\epsilon_2 B_0 S^z_2 +\alpha S^x_1 S^x_2,
\end{equation}
which is somewhat reminiscent of the usual XXZ Hamiltonian widely
used in statistical mechanics. Spin Hamiltonians have been
considered in connection to quantum computation proposals using
NMR \cite{n403jaj2000,prl86ysw2001} (see also \cite{roadmap}) and
cold atoms in optical lattices
\cite{n415mg2002,prl81dj1998,pra69vm2004}. Entanglement in spin
chains at finite temperature was considered for example in
\cite{pra64dg2001,prl87mca2001}. For simplicity we shall assume
$\epsilon_1 B_0=\epsilon_2 B_0=1$. This very simple case already
shows a rich behavior depending on the parameter $\alpha$ and the
spin magnitudes $s_1$ and $s_2$.

The article is organized as follows: in section I we consider
$s_1=s_2=1/2$, which corresponds to a pair of qubits. In section
II we let $s_1\to \infty$ to simulate an environment and study its
ability to induce decoherence upon spin $s_2$. Section 3 deals
with the relation between chaos and decoherence and we conclude in
section IV.

\section{Two qubits ($s_1=s_2=1/2$)}
Let us first consider the case $s_1=s_2=1/2$. The Hamiltonian in
this case is a $4\times 4$ matrix that can easily be diagonalized.
Its eigenvectors, written in the standard $st:=\{|11\rangle,
|10\rangle, |01\rangle, |00\rangle\}$ basis are
\begin{equation}
\begin{array}{cc}
v_1=\left(
\begin{array}{c}
0 \\1\\1\\0
\end{array}
\right), \quad v_2=\left(
\begin{array}{c}
0 \\-1\\1\\0
\end{array}
\right), \\ v_3=\left(
\begin{array}{c}
4(1+\beta)/\alpha \\0\\0\\1
\end{array}
\right), \quad v_4=\left(
\begin{array}{c}
4(1-\beta)/\alpha \\0\\0\\1
\end{array}
\right), \end{array}
\end{equation}
with respective eigenvalues
\begin{equation}\label{eigen}
e_1=\frac{\alpha}{4},e_2=-\frac{\alpha}{4},e_3=\beta,e_4=-\beta,
\end{equation}
where
\begin{equation}
\beta=\frac{\sqrt{\alpha^2+16}}{4}.
\end{equation}

In order to study dynamical entanglement generation, we use
several different initially separable configurations. At $t=0$,
the spins $s_1$ and $s_2$ are in the states
\begin{equation}
|z_i\rangle=\frac{|0\rangle+z_i|1\rangle}{\sqrt{1+|z_i|^2}}, \quad
i=1,2
\end{equation}
and, after some symmetry considerations, we have decided to
investigate the following values of the parameters:
\begin{equation}
\begin{array}{cc}
&a) z_1=z_2=0,  \quad b) z_1=1,z_2=0\\
&c) z_1\to \infty, z_2=0,  \quad d) z_1=0,z_2=1\\
&e) z_1=0,z_2={\rm i},  \quad f) z_1=z_2=1\\
&g) z_1=1,z_2={\rm i},  \quad h) z_1=z_2={\rm i}
\end{array}
\end{equation}
Note that $z_1=\pm 1$ and $z_1=\pm{\rm i}$ correspond to
eigenstates of $S^x_1$ and $S^y_1$, respectively, and the same
holds for spin $s_2$.

A measure of entanglement exists for two spins: the entropy of
formation, which measures ``the amount of entanglement necessary
to create the entangled state" \cite{pra67tcw2003}. If the system
is initially separable, which is the case we consider here, the
entropy of formation is equal to the von-Neumann entropy. For each
one of the chosen initial conditions we followed the time
evolution of this quantity and of linear entropy. Both measures
have approximately the same form, with maxima and minima at the
same points, even tough they are not proportional to each other.
We have also observed that $\delta\leq\delta_N$ for all chosen
initial conditions, but this is not a general property, since the
opposite occurs in section III. In Fig. 1a we see the case $a)$,
which is very similar to cases $b)-e)$, all in which at least one
of the spins is in the $|0\rangle$ state. Conditions $f)-h)$ are
also quite similar, all in which $|z_1|=|z_2|=1$. Condition $f)$
can be seen in Fig. 1b. All time scales present in the system can
be related to the eigenvalues (\ref{eigen}), and are therefore
determined by $\alpha$.

\section{A qubit and an environment ($s_1\to \infty, s_2=1/2$)}

We now consider $s_2=1/2$ and $s_1=200$. Such a huge value of spin
is not intended to be realistic, but only to simulate an
environment for dissipation, when infinitely many degrees of
freedom are required (see \cite{rmp75whz2003} and references
therein). We consider spin $s_1$ to be, at $t=0$, in three
different conditions: in a coherent state \cite{klauder1985}
\begin{equation}\label{cs}
|z_1\rangle=\sum_{m=-s_1}^{s_1}\sqrt{\left(
\begin{array}{c}
2s_1 \\
s_1+m
\end{array}
\right) }\frac{z^{s_1+m}}{(1+|z|^{2})^{s_1}}| m\rangle,
\end{equation}
in a non-localized state
\begin{equation}\label{nonloc}
|\psi\rangle=(2s_1+1)^{-\frac{1}{2}}\sum_{m_1=-s_1}^{s_1}
|m_1\rangle
\end{equation}
and in a thermal mixture
\begin{equation}\label{therm}
\rho_1=\frac{1}{N}\sum_{m_1=-s_1}^{s_1}
e^{-m_1/T}|m_1\rangle\langle m_1|,
\end{equation}
where $N$ is a normalization factor (the partition function)

Let us first consider both spins to be initially  in coherent
states with $z_1=z_2=0$. The linear entropy is shown in Fig. 2
(the von-Neumann entropy is indistinguishable from it in this
scale). We see that the entanglement reaches its maximum value
very fast, and remains saturated for some time, but a strong and
localized recoherence takes place in a nearly periodic way. As
already observed in \cite{prl80kf1998}, such recoherences are
related to the mean value of the $z$-component of the spin,
$\langle S^z_2\rangle$, or equivalently to how close is the spin
to the north pole of the Bloch sphere \cite{Bloch}. In other
words, recoherence is an effect of the compactness of the phase
space.

In Fig. 3a we see the results for $z_1=0, z_2=1$. Even though the
value of $z_1$ is the same as in the previous case, the
entanglement is completely different. The linear entropy is now
less than the von-Neumann entropy, but they have essentially the
same information. Entanglement increases with time, but much
slower than the previous case. Instead of raising immediately to
its maximum value, it increases by jumps, alternating fast
increments with plateaus.

We also consider $z_1=z_2=1$, shown in Fig. 3b. The behavior of
the entropies is quite similar to the previous case, but in a very
different scale: entanglement is nearly a hundred times smaller in
this case, i.e., the system is practically entanglement-free. It
is clear that the entanglement rate strongly depends on the
initial state of the environment, even if this is a coherent
state.

Finally, we have considered the environment to be initially in the
non-localized state (\ref{nonloc}) and in the thermal mixture
(\ref{therm}). In all cases we have used $z_2=1$, and in Fig. 4 we
show the behavior of the linear entropy. The qualitative behavior
is the same for all initial states (in the thermal mixture we have
used $T=s_1/10$), but the ability to induce decoherence is
considerably greater for the coherent state.

\section{Entanglement and chaos ($s_1=s_2\to\infty$)}
We now consider both spins to be large (semiclassical), namely we
use $s_1=s_2=15$. For any spin Hamiltonian it is possible to
obtain a semiclassical dynamics using $su(2)$ coherent states
(\ref{cs}), in which the quantum dynamics is then approximated by
Hamilton equations of motion \cite{klauder1985}
\begin{equation}\label{semi}
\dot{z}=-{\rm i}\frac{\left( 1+|z|^{2}\right) ^{2}}{2s\hbar
}\frac{\partial \mathcal{H}}{\partial z^{\ast }},
\end{equation}
where $\mathcal{H}=\langle z|H|z\rangle$ plays the part of a
classical Hamiltonian. In our case this is given by
\cite{prl61wmz1988}
\begin{equation}
\mathcal{H}=\frac{1}{2}(A_1+A_2-31)+\frac{\alpha}{4}
q_1q_2\sqrt{(4s_1-A_1)(4s_2-A_2)},
\end{equation}
where
\begin{equation}
A_i=q_i^2+p_i^2
\end{equation}
and the canonical coordinates are defined by
\begin{equation}
\frac{q_j+{\rm i}p_j}{\sqrt{4s_j}}=\frac{z_j}{\sqrt{1+|z_j|^2}},
\quad j=1,2.
\end{equation}

In Fig. 5 we see a Poincar\'{e} section showing the $(q_1,p_1)$
coordinates at $p_2=0$. We use the marked points as initial
conditions for our coherent states to be evolved. The trajectory
marked with a triangle is chaotic, the one marked with a square is
regular and the circle denotes a periodic orbit. Note that we do
not evolve the state semiclassically: a semiclassical evolution of
the kind (\ref{semi}) would preserve, by construction, the
coherence of the initial state, and thus produce no entanglement.

In Fig. 6 we see the linear entropy as a function of time. The
rate of entanglement for short times is very similar for all
initial conditions, contrary to what is observed for example in
the Jaynes-Cummings system \cite{prl80kf1998}. We also note that
the maximal value of $\delta$ is higher for the periodic orbit
than for the more generic regular trajectory. Numerical analysis
indicate that both the short-time entanglement rate and the
maximum entanglement value are related not to the classical
dynamics but to strictly quantum properties of the wave packet,
such as energy uncertainty.

There is however a strong difference between the regular and
chaotic initial conditions: the former present strong recoherences
that are not present in the latter. As already observed in the
previous section these recoherences are related to the mean value
of the $z$-component of the spin, but in this case the relevant
spin is $s_1$, as we see in Fig. 7. Note that coherent states have
localized Husimi distributions in phase space and thus some more
general, less localized state, would be less likely to display
recoherences.

Again the entanglement measures are qualitatively very similar,
even tough they are not proportional, but in this case the
von-Neumann entropy is less than the linear entropy at all times
and for all initial conditions. We show this only for the periodic
orbit in Fig. 8.

\section{Summary}

We have studied dynamical entanglement generation for a pair of
interacting spins in a magnetic field. The case of two qubits is
rather simple, since it allows an analytic treatment up to a point
and is useful as a background against which we can compare the
other cases. We noted for example that for all initial conditions
$\delta\leq \delta_N$. The case of a qubit interacting with an
environment had a much more rich variety. We have seen that the
decohering power of the environment strongly depends on its
initial state. A comparison was made between different coherent
states and between a coherent state, a non-localized state and a
thermal mixture.

In the case of two semiclassical spins we have analyzed the
influence of chaos upon the entanglement process. We saw that
chaotic initial conditions presented no recoherences, while
regular ones do so. We also noted that in this case
$\delta\geq\delta_N$ for all initial conditions, contrary to the
low dimensional case of two qubits. In all cases the recoherences
present in time evolution were related to the compactness of spin
phase space and localization of coherent states. It would be
interesting to investigate other measures of entanglement,
specially for the case of initially mixed systems. Such work is
underway.

\bibliography{references}
\bibliographystyle{apsrev}

\newpage

\begin{figure}[htb]
\includegraphics[scale=0.5,angle=-90]{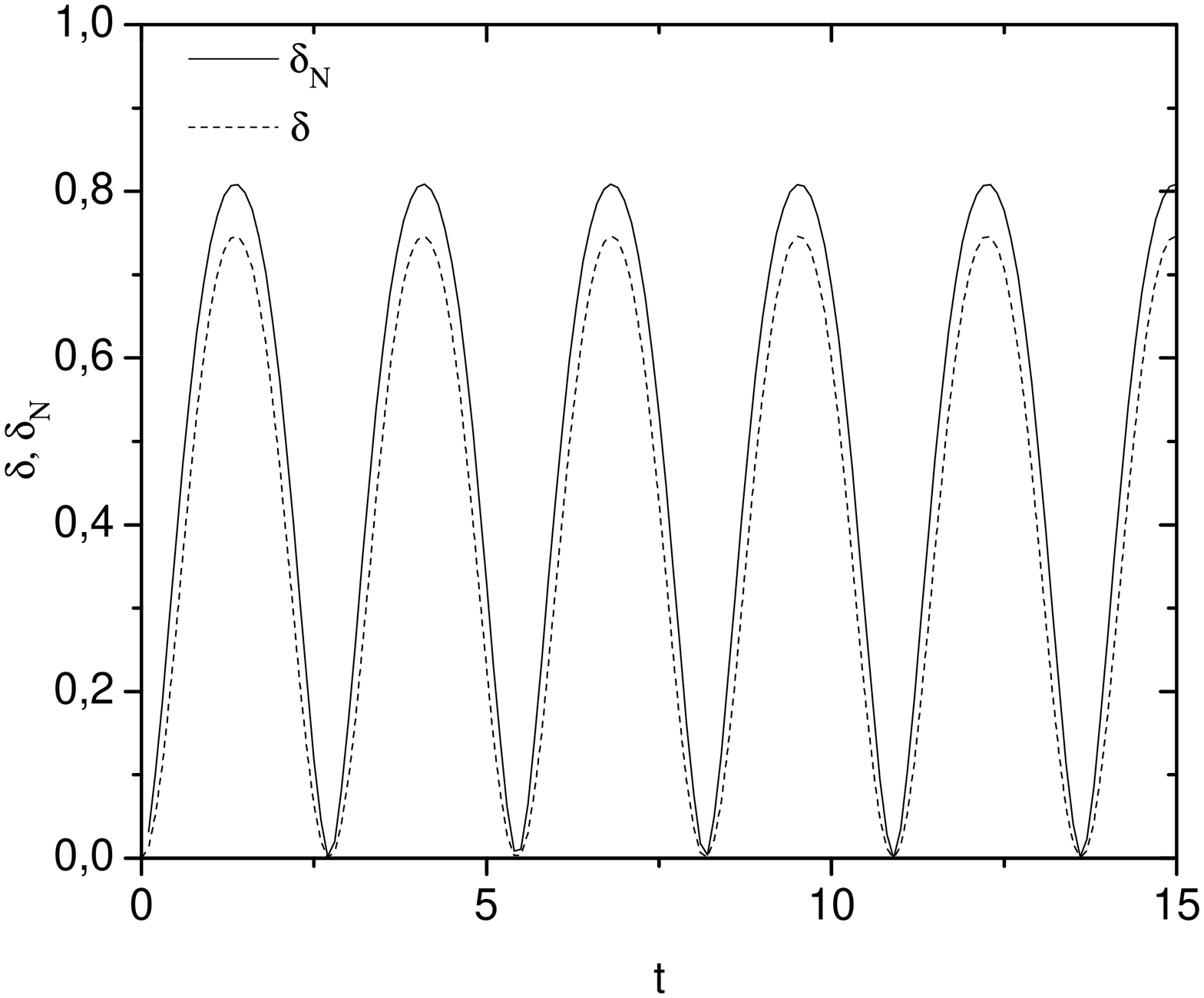}
\includegraphics[scale=0.5,angle=-90]{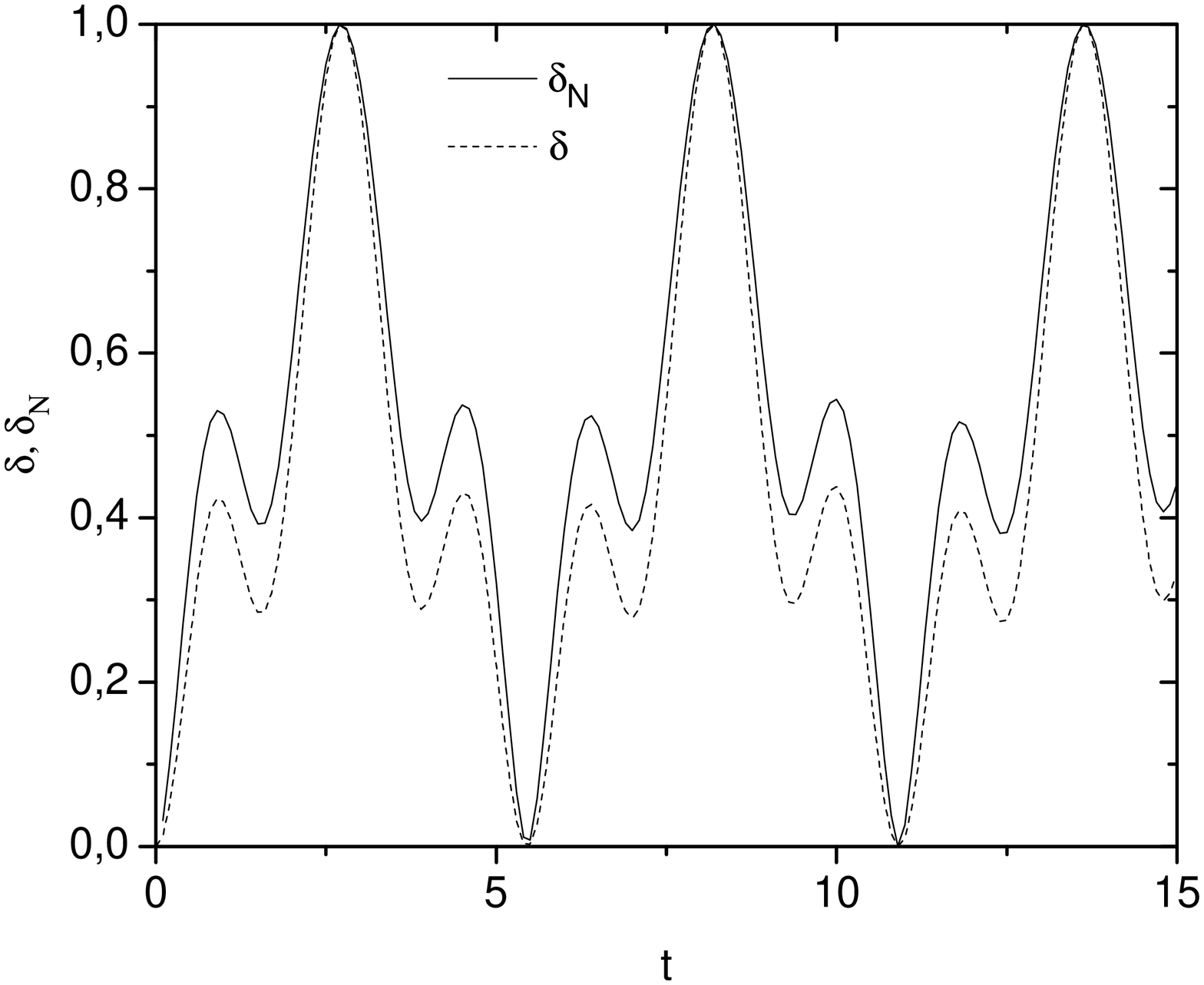}
\caption{($s_1=s_2=1/2$) Linear (dashed line) and von-Neumann
(solid line) entropies for $z_1=z_2=0$ (upper panel) and for
$z_1=z_2=1$ (lower panel). We have observed that $\delta_N\geq
\delta$ for all times and all initial conditions. }
\end{figure}

\begin{figure}[htb]
\includegraphics[scale=0.5,angle=-90]{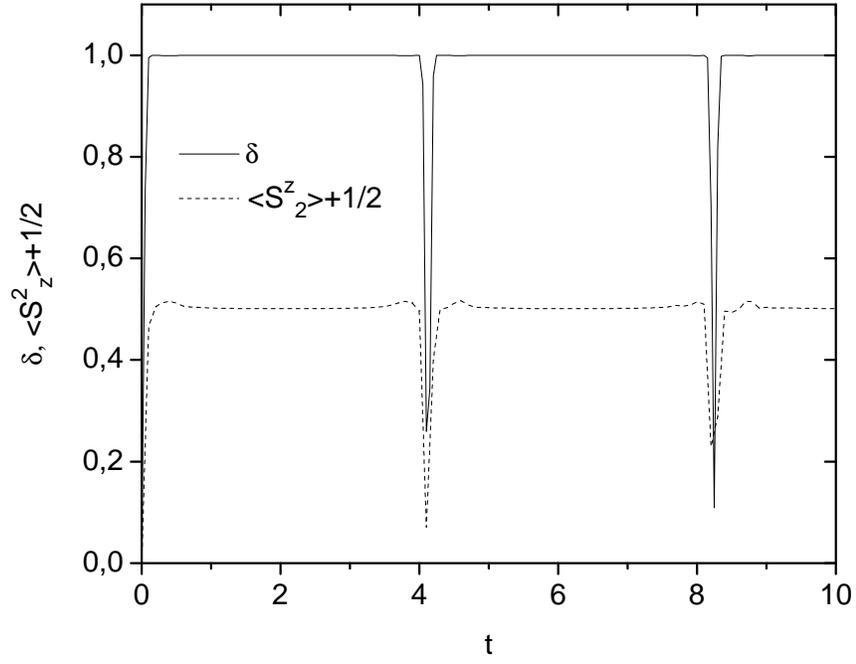}
\caption{($s_1=200$, $s_2=1/2$) Linear entropy (solid line) and
angular momentum (dashed line) as functions of time for
$z_1=z_2=0$. We see that $\delta$ saturates very fast, but strong
recoherences appear, related to $\langle S_z^2\rangle$.}
\end{figure}

\begin{figure}[htb]
\includegraphics[scale=0.5,angle=-90]{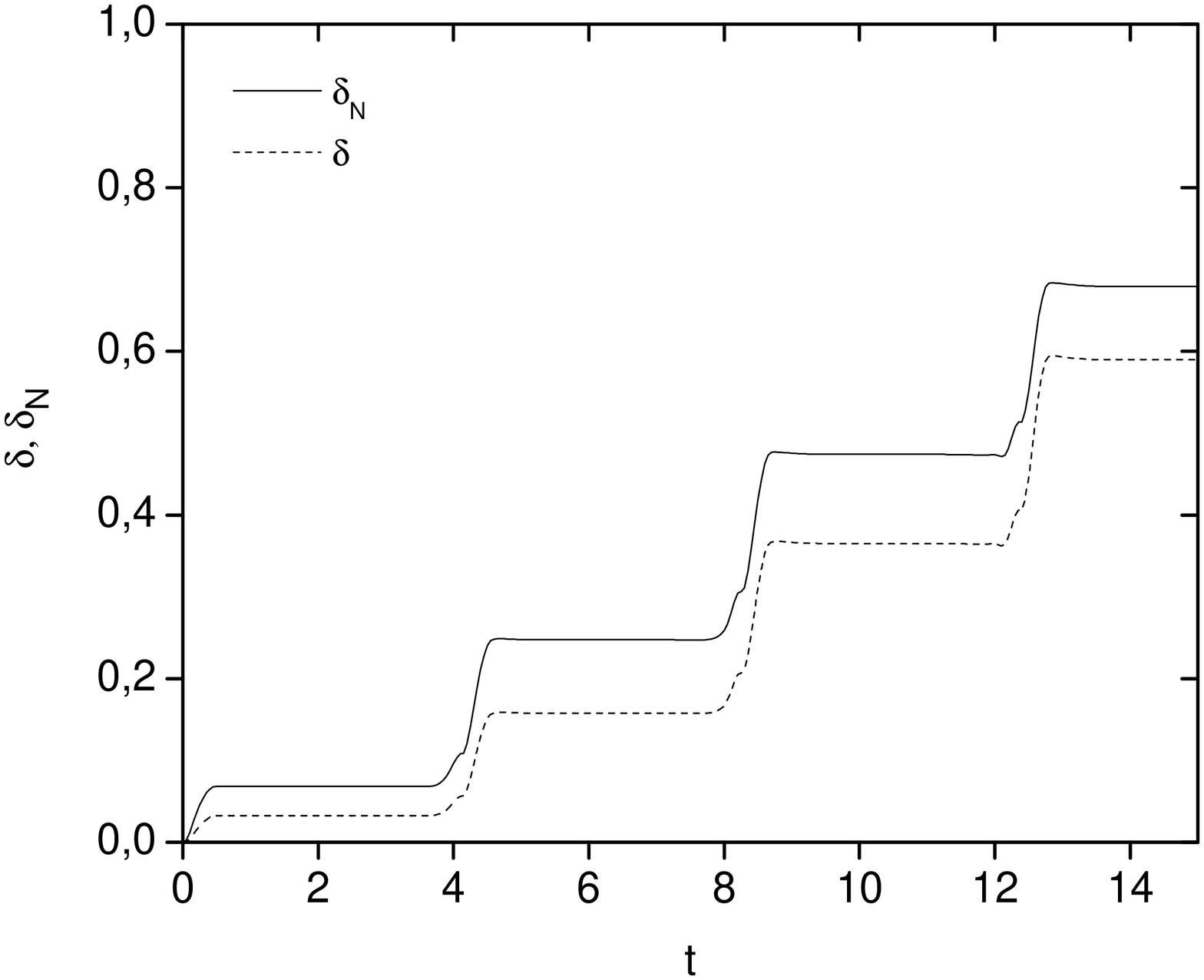}
\includegraphics[scale=0.5,angle=-90]{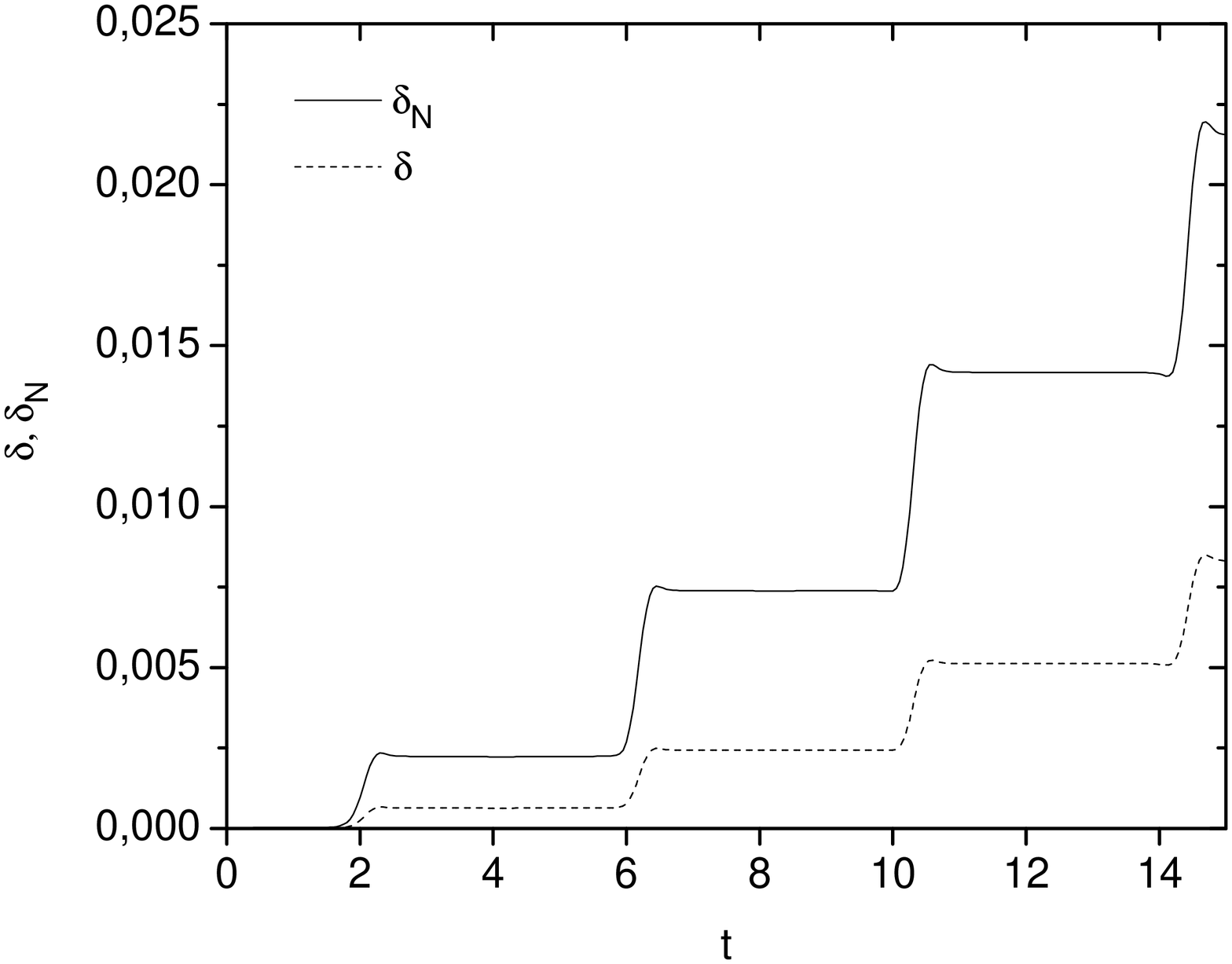}
\caption{($s_1=200,s_2=1/2$) Linear (dashed line) and von-Neumann
(solid line) entropies for $z_1=0, z_2=1$ (upper panel) and for
$z_1=z_2=1$ (lower panel). In both cases we see plateaus followed
by jumps, but there is a large difference in magnitude.}
\end{figure}

\begin{figure}[htb]
\includegraphics[scale=0.5,angle=-90]{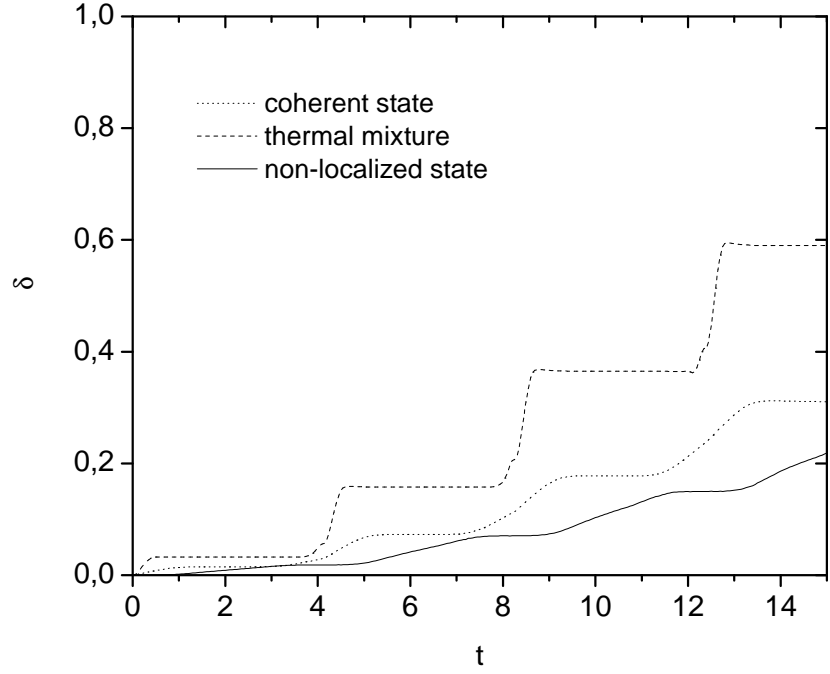}
\caption{($s_1=200,s_2=1/2$) Linear entropy evolution, for the
environment initially in a coherent state (dashed line), a thermal
mixture (dotted line) and a non-localized state (solid line).}
\end{figure}

\begin{figure}[htb]
\includegraphics[scale=0.5,angle=-90]{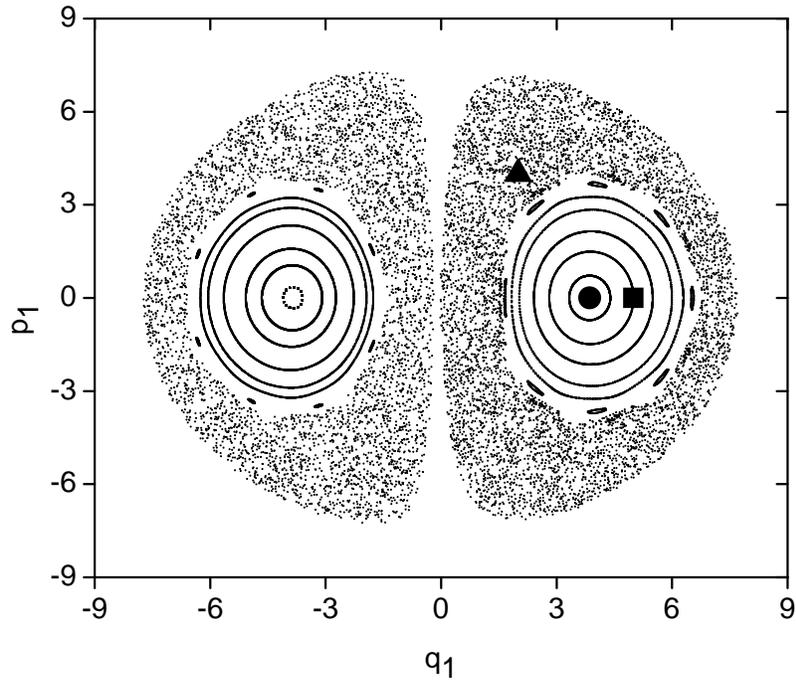}
\caption{ Poincar{\'{e}} section of the classical Hamiltonian. We
use the marked points as initial conditions for our coherent
states to be evolved.}
\end{figure}

\begin{figure}[htb]
\includegraphics[scale=0.5,angle=-90]{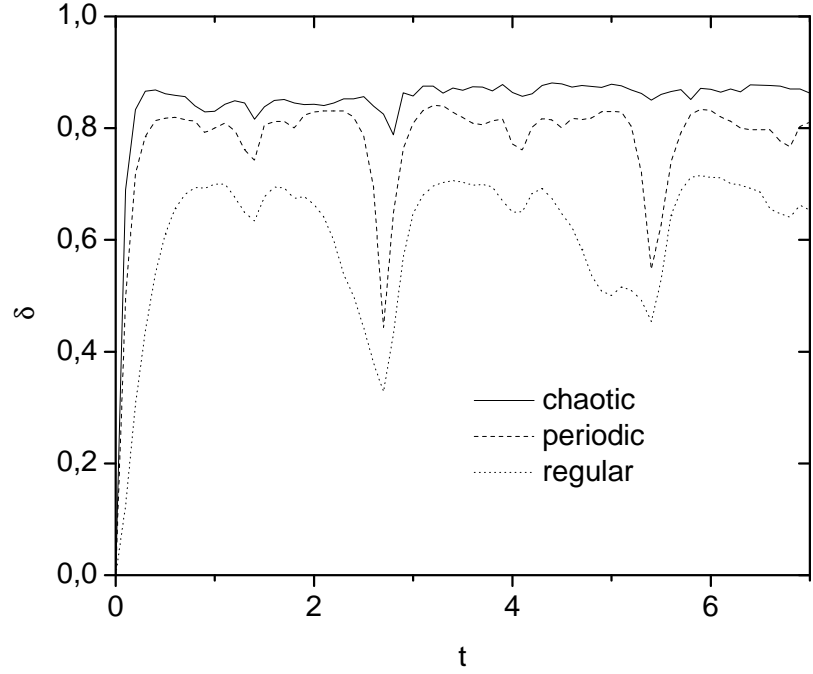}
\caption{($s_1=s_2=15$) Linear entropy $\delta(t)$ for three
different initial conditions. The regular cases present strong
recoherences, while these are suppressed in the chaotic case.}
\end{figure}

\begin{figure}[htb]
\includegraphics[scale=0.5,angle=-90]{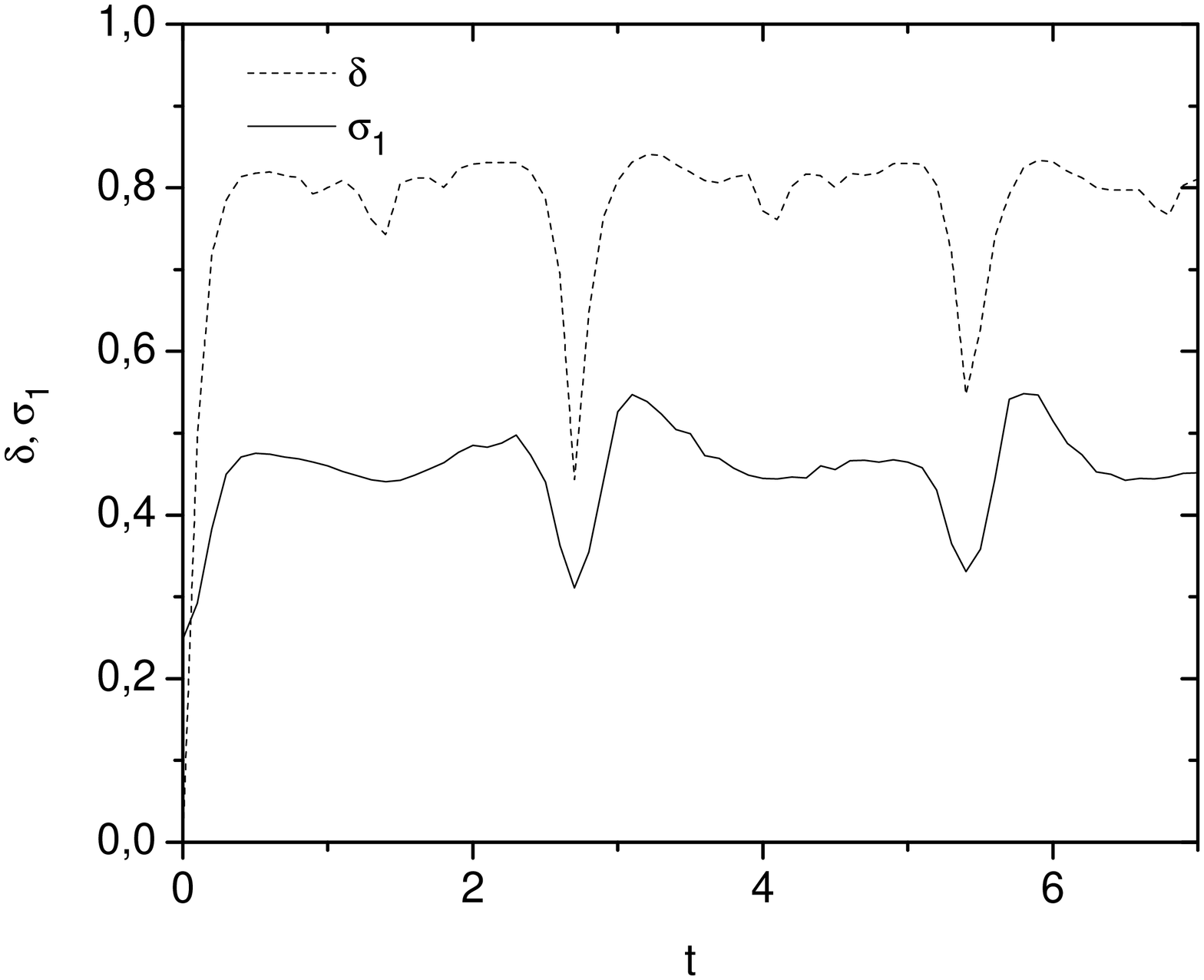}
\caption{($s_1=s_2=15$) Linear entropy and angular momentum
$\sigma_1=\frac{\langle S^z_1 \rangle+s_1}{2s_1}$ for the periodic
initial condition. Recoherences are related to the compactness of
phase space.}
\end{figure}

\begin{figure}[htb]
\includegraphics[scale=0.5,angle=-90]{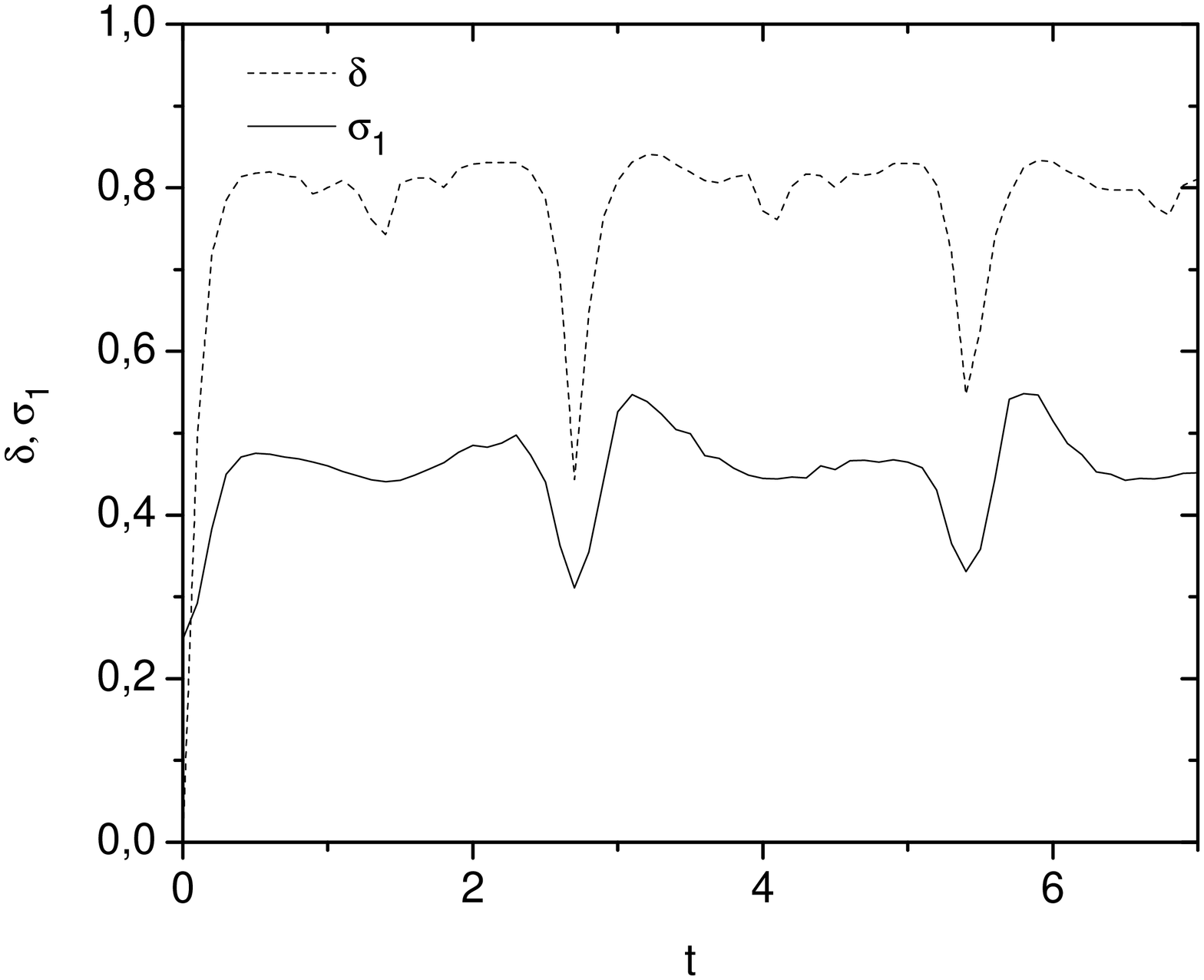}
\caption{($s_1=s_2=15$) Linear (dashed line) and von-Neumann
(solid line) entropies for the periodic initial condition. In
contrast to section I, now $\delta\geq\delta_N$ for all times and
all initial conditions.}
\end{figure}

\end{document}